\documentclass{sig-alternate-2013}

\permission{Copyright is held by the International World Wide Web Conference Committee (IW3C2). IW3C2 reserves the right to provide a hyperlink to the author's site if the Material is used in electronic media.}
\conferenceinfo{WWW'16 Companion,}{April 11--15, 2016, Montr\'eal, Qu\'ebec, Canada.} 
\copyrightetc{ACM \the\acmcopyr}
\crdata{978-1-4503-4144-8/16/04. \\
http://dx.doi.org/10.1145/2872518.2890571 }

\usepackage{footnote}
\usepackage{hyperref}

\begin{document}


%
%
%

%

\title{Question Answering on Linked Data: Challenges and Future Directions}
%
%
%
%
%

\numberofauthors{7} 

\author{
\alignauthor
Saeedeh Shekarpour\\
\affaddr{Knoesis Research Center, USA}\\
\email{saeedeh@knoesis.org} 
\and
\alignauthor
Kemele M. Endris\\
\affaddr{University of Bonn, Germany}\\
\email{endris@cs.uni-bonn.de}
\and
\alignauthor
Ashwini Jaya Kumar\\
\affaddr{Fraunhofer IAIS, Germany}\\
\email{ashwinijk18@gmail.com}
\and
\alignauthor
Denis Lukovnikov\\
\affaddr{University of Bonn, Germany}\\
\email{lukovnik@cs.uni-bonn.de} 
\alignauthor
Kuldeep Singh\\
\affaddr{Fraunhofer IAIS, Germany}\\
\email{kskuldeepvit@gmail.com}
\alignauthor
Harsh Thakkar\\
\affaddr{University of Bonn, Germany}\\
\email{thakkar@cs.uni-bonn.de}
\and
\alignauthor
Christoph Lange\\
\affaddr{University of Bonn / Fraunhofer IAIS, Germany}\\
\email{langec@cs.uni-bonn.de}
}
\maketitle

\begin{abstract}
Question Answering (QA) systems are becoming the inspiring model for the future of search engines.
While, recently, datasets underlying QA systems have been promoted from unstructured datasets to structured datasets with semantically highly enriched metadata, question answering systems are still facing serious challenges and are therefore not meeting users' expectations. 
This paper provides an exhaustive insight of challenges known so far for building QA systems, with a special focus on employing structured data (i.e. knowledge graphs).
It thus helps researchers to easily spot gaps to fill with their future research agendas.
\end{abstract}

%
%
%

%
%

%
%


\keywords{Question Answering System, Research Challenge, Speech Interface, Query Understanding, Data Quality, Distributed and Heterogeneous Datasets, Interoperability of Components.}

\section{Introduction}

The Web of Data is growing enormously (currently more than 84 billion triples\footnote{observed on 14 October 2015 at \url{http://stats.lod2.eu/}}).
This figure comprises both structured and unstructured data.
Still, taking advantage of this rapidly growing amount of data is challenging. 
Traditional information retrieval approaches based on keyword search are user-friendly but fail to exploit the internal structures of data due to their bag-of-words semantics.
For searching information on the Data Web we need similar user friendly approaches, i.e. keyword-based interfaces, which rely on the internal structure of the data. 

Question Answering (QA) is a specialized form of information retrieval. 
A Question Answering system retrieves exact answers to questions posed in natural language by the user. 
While, recently, datasets underlying QA systems have been promoted from unstructured datasets to structured datasets with semantically highly enriched metadata, question answering systems are still facing serious challenges and are therefore not meeting users' expectations. 

Question Answering systems consists of components that can be studied and evolved independently.
These components include (1) an input interface for obtaining a query, (2) components for understanding, interpreting, disambiguating and parsing the query, (3) components accessing and processing the datasets employed (facing issues such as heterogeneity, quality and indexing); thus, there are also issues of (4) interoperability among different interacting components.
In the following, we elaborately discuss challenges related to each aspect and consider future research directions.
We close with a conclusion and a roadmap for future work.


\section{Challenges}
In this section we present question answering challenges from four different aspects namely, (i) Speech-based interface challenge, (ii) query understanding, interpreting, disambiguating and parsing challenges, (iii) data-oriented challenges (iv) interoperability of QA components challenge.  

\subsection{Speech Interface}

Interfacing speech to QA systems has become a focus of research for a long time. But, the main focus of research effort so far has been spent on interfacing speech to IR-based QA systems ~\cite{VAQAreview,sanchis2006spoken,huspeechqoogle}, and much less on interfacing speech input to QA systems based on KGs (knowledge graphs). Typical state-of-the-art IR approaches integrate a speech recognition (SR) unit directly with the QA system. An effort beyond merely interfacing the two units is required to enable natural conversation in question answering system for both IR and KG methods.

An SR system mainly consists of an acoustic model and a language model, where the main objective is to decode what is uttered by the user. In contrast, a general IR based QA system comprises question processing (to extract the query from the input and to determine the answer type), passage retrieval, document retrieval, passage extraction, and finally answer selection depending on the relatedness of the named entities found to the question keyword. The accuracy of recognizing spoken words has a vital influence on the success of the whole QA process. Ex: if `Jamshedpur' (a city in India) is recognised as `game shed poor' (articulation style and duration of utterance is the key difference), then the whole QA process is altered. The city name which constitute the important answer type is not recognised by the QA system. This can be avoided if there is a rich dataset to train a recogniser but it is not possible to have acoustic training data for an open-domain.  Hence speech recognisers are usually built for a specified domain. The same applies for QA systems, developing an open-domain QA is a challenge.

With the evolution of neural network based methods for speech recognition, the whole conventional approach to speech recognition has changed. Generally, acoustic model and language model were built as two independent units. The development of single neural network architecture to transcribe an audio input is a breakthrough in the speech recognition research ~\cite{DBLP:journals/corr/HannunCCCDEPSSCN14, graves2014towards, amodei2015deep}. The recognition accuracy has been tested for a character level transcription and it is indicated that a word/sentence level transcription can be made with the same architecture. In this type of single neural network based speech recognition, language model is applied at the output of speech recogniser. Following the same methodology, it is possible to build an end-to-end speech interfaced QA system with deep neural networks. Current research direction is towards exploring the interface of speech to knowledge graph using deep neural networks.

\subsection{Understanding Questions}
\subsubsection{Discussion}
In the case of full-fledged QA over structured data, for example over a knowledge base (KB) such as Freebase~\cite{bollacker2008freebase}, the question must be translated into a logical representation that conveys its meaning in terms of entities, relations, types as well as logical operators. Simpler forms of QA can also be achieved in other ways, however, approaches without formal translation can not express certain constraints (e.g. comparison).
The task of translating from NL to a logical form (semantic parsing (SP)) is characterized by the mismatch between natural language (NL) and knowledge base (KB).
The semantic parsing problem can be divided into two parts: (1) determining KB constituents mentioned in the NL expression and (2) determining how these constituents should be arranged in a logical structure.
The mismatch between NL and KB brings several problems.
One problem is Entity Linking (EL), recognizing parts of NL input that refer to an entity (NER) and determining which named entities are meant by that part (disambiguation).
A central challenge in EL is how to take into account the context of an entity mention in order to find the correct meaning (disambiguation).
Another challenge is finding an optimal set of suitable candidates for a mention, where the \emph{lexicon} (mapping between words/phrases and entities) plays an important role. A problem bordering both disambiguation and candidate generation is the large number of entities a word can refer to (e.g. the thousands of possible ``\textit{John}'''s when confronted with ``\textit{John starred in 1984}'').

Another problem is relation detection and classification.
Given an NL phrase, we want to determine which KB relation is implied by the phrase. Sometimes, the relation is explicitly denoted by a NL constituent, for example verb-mediated statements (e.g. ``$X$ married $Y$''), in which case a lexicon can help a lot to solve the problem.
However, in general, a lexicon-based approach is not sufficient.
Sometimes there are no relation-specific words in the sentence.
Sometimes prepositions are used, for example ``\textit{works by Repin}'' or ``\textit{cars from Germany}'' and sometimes the semantics of the relations and the entities/types they connect are lexicalized as one, for example, ``\textit{Russian chemists}'' or ``\textit{Tolstoy plays}''.
Such cases require context-based inference, taking into account the semantics of the entities that would be connected by the to-be-determined relation (which in turn is related to parsing).


Merely linking entities and recognizing the relations is not sufficient to produce a logical representation that can be used to query a data source.
The remaining problem is to determine the overall logical structure of the NL input.
This problem becomes difficult for longer, more complex sentences, where different linguistic phenomena, such as coordination and co-reference, must be handled.
Formal grammars, such as CCG~\cite{ccg}, can help to parse NL input.
CCG in particular is well-suited for semantic parsing because of its transparent interface between syntactic structure and underlying semantic form.
One problem with grammar-based semantic parsers is their rigidity, which is not well-suited for incomplete input as often found in real-world QA scenarios.
Some works have explored learning relaxed grammars~\cite{relaxedccg} to handle such input.

The straightforward way of training semantic parsers requires training data consisting of NL sentences annotated with the corresponding logical representation, which are very cumbersome to obtain.
Recent works have explored different ways to reduce the annotation effort in order to bypass this challenge.
One proposed way is to train on question-answer pairs instead~\cite{parasempre}. 
Another way is to automatically generate training data from the KB 
and/or from entity-linked corpora~\cite{reddy} 
 (e.g. ClueWeb). 
Training with paraphrasing corpora~\cite{parasempre} 
is another technique explored in several works to improve the range of expressions the system will be able to cover.


\subsubsection{Future directions}
Recently, impressive advances in different tasks in Artificial Intelligence have been achieved using deep learning techniques.
Embedding-based language models, such as Word2Vec~\cite{mikolov1,mikolov2} and GloVe~\cite{glove}, have helped to improve performance in many NLP tasks. 
One of the most interesting and the most promising future directions for semantic parsing and question answering is further exploration of deep learning techniques in their context.


Using deep learning to better understand questions can be done by using (possibly custom-trained) word (, word sense and entity) embeddings, which capture their syntactic and semantic properties, as features to improve existing workflows.
However, a ``deeper'' approach would be to also devise new models that provide the machine with more freedom to figure out how to accomplish the task.
An excellent and very recent example in NLP is the Dynamic Memory Network (DMN~\cite{dmn}), that does not use any manually engineered features or problem-tailored models, and yet achieves state-of-the-art performance on all tested tasks, which are disjoint enough to leave one impressed (POS tagging, co-reference resolution, sentiment analysis and Question Answering on the bAbI dataset).
The DMN is one of the works focusing on attention and memory in deep learning that enables the neural network to reason more freely.
We share the belief that the investigation and application of more advanced deep learning models (such as DMN and NTM~\cite{ntm}) could yield impressive results for different tasks in AI, including question answering.

Recursive, convolutional (CNN) and recurrent (RNN) neural networks are widely used in recent neural network-based approaches.
Convolutional Neural Networks (CNN), a special case of recursive NNs are well-explored for computer vision.
Recursive NNs have also been applied for parsing and sentiment analysis. 
RNNs produce state-of-the-art results in speech processing as well as in NLP because of their natural vigor for processing variable-length sequences.
They have been applied for machine translation (SMT)~\cite{smtrnn}, language generation (NLG)~\cite{nlgrnn}, language modeling and more and are also fundamental for the success of the DMN and the NTM.

Even though the DMN has not yet been applied to our task of structured QA, some recent works, such as the relatively simple embedding-based work of Bordes et al. ~\cite{bordes2014question} (which outperformed ParaSempre~\cite{parasempre} on \textsc{WebQuestions}) and the SMT-like SP approach of~\cite{dong2016language} seem to acknowledge the promise of neural approaches with embeddings.

An additional interesting direction is the investigation of joint models for the sub-problems involved in question interpretation (EL, co-reference resolution, parsing, \dots).
Many tasks in NLP depend on each other to some degree, motivating the investigation of efficient approaches to make the decisions for those tasks jointly.
For example, co-reference resolution and EL can benefit from each other as entity information from a KB can serve as quite powerful features for co-reference resolution and co-reference resolution in turn can improve EL as it transfers KB features to phrases where anaphora refer to entities.
Factor Graphs (and Markov Networks) are by nature very well-suited for explicit joint models (e.g.~\cite{jointriedel}).
However, a more internal kind of joint inference could also be achieved within a neural architecture (e.g. the DMN).

However, it is worth noting that training advanced neural models and explicit joint models can be a difficult task because of the large number of training parameters and co-dependence of these parameters. Deep learning typically relies on the availability of large datasets. However, the whole task to be solved can be divided in two parts, one focusing on representation learning, which can accomplished in an unsupervised setting (with large amounts of data) and the second part relying on and possibly fine-tuning the representations obtained in the first part in a supervised training setting (requiring annotated task-specific data). For explicit joint models, data capturing the dependence between different task-specific parts of the models (e.g. annotated for both EL and co-reference) are required and the efficient training of such models is a very relevant current topic of investigation.

The concluding thought is that the further investigation of language~\cite{mikolov1,mikolov2,glove,sensembed} and knowledge modeling~\cite{tatec,transe,riedel2013,ntn,trescal} and powerful deep neural architectures with self-regulating abilities (attention, memory) as well as implicit or explicit joint models will continue to push the state of the art in QA.
Well-designed deep neural architectures, given proper supervision and powerful input models, have the potential to learn to solve many different NLU problems robustly with minimal customizations, eliminating the need for carefully engineered features, strict formalisms to extract complex structures or pipelines arranging problem-tailored algorithms.
We believe that these lines of research in QA could be the next yellow brick~\cite{yellowbrick} in the road to true AI, which has fascinated humanity since the ancient tales of Talos and Yan Shi's mechanical men.


\subsection{Data challenges}
\subsubsection{Indexing Heterogeneous Datasets}
A typical QA system is empirically only as good as the performance of its indexing module~\cite{dong2007indexing}.
The performance of indexing serves as an upper bound to the overall output of the QA system, since it can process only as much data as is being presented/served to it from the indices.
The precision and recall of the system may be good, but if all or most of the top relevant documents are not indexed in the system, the system performance suffers and hence does the end user.

Many researchers have compared effectiveness across a variety of indexing techniques.
Their studies show improvement if multiple techniques were combined compared to any single individual indexing technique~\cite{rajashekar1995combining}.
In the present scenario, information retrieval systems are carefully tailored and optimized to deliver highly accurate results for specific tasks.
Over the years, efforts of developing such task specific systems have been diversified based on a variety of factors discussed in the following. 

Based on the type of the data and the application setting, a wide range of indexing techniques are deployed.
They can broadly be categorized into three categories based on the format and type of data indexed, namely: structured (e.g. RDF, SQL, etc.), semi-structured (e.g. HTML, XML, JSON, CSV, etc.) and/or unstructured data (e.g. text dumps).
They are further distinguished by the type of technique they use for indexing and/or also by the type of queries that a particular technique can address.
The different techniques inherently make use of a wide spectrum of underlying fundamental data structures in order to achieve the desirable result.

Most of the systems dealing with unstructured or semi-structured data make use of inverted indices and lists for indexing.
For structured datasets, a variety of data structures such as AVL trees, B-Trees, sparse indices, IR trees, etc., have been developed in the past decades.
Many systems combine two or more data structures to maintain different indices for different data attributes.
We present a short survey of indexing platforms and data structures used in a wide range of QA systems in table~\ref{comp_table_indexing}. 

\begin{table*}[]
\centering
\resizebox{\textwidth}{!}{%
\begin{tabular}{|p{0.3\textwidth}|p{0.35\textwidth}|p{0.35\textwidth}|}
\hline
\textbf{System} & \textbf{Data structure used} & \textbf{Platform used for indexing} \\ \hline
\textbf{SWSE/YARS2 \cite{Harth2007}} & Sparse, Inverted Indices for RDF quads & Lucene \\ \hline
\textbf{Sindice \cite{Oren}} & Inverted Index and On-disk persistent storage & Solr \\ \hline
\textbf{Sina \cite{shekarpour2015sina}} & Bitmap index on RDF quads (in total 5 indices are maintained: 2 full RDF quad indices, 3 partial RDF quad indices) & OpenLink Virtuoso \\ \hline
\textbf{HAWK \cite{Usbeck} } & *N/A & *N/A \\ \hline
\textbf{TBSL \cite{unger2012template}} & Inverted Index & Solr \\ \hline
\textbf{Ephyra \cite{schlaefer2007semantic}} & Inverted index & Lemur-Indri \\ \hline
\textbf{PowerAqua \cite{lopez2009cross}} & Inverted index & Lucene (two indices are prepared taxonomically) \\ \hline
\textbf{AquaLog \cite{lopez2005aqualog}} & *N/A & GATE MIMIR possibly with Lucene \\ \hline
\textbf{Sig.ma \cite{tummarello2010sig} } & Inverted Index and On-disk persistent storage & Solr \\ \hline
\textbf{QUADS \cite{Yang2014}} & Inverted index & Lucene \\ \hline
\textbf{MAYA \cite{Kim2001}} & (key, value) pairs & Traditional index with RDBMS \\ \hline
\textbf{ESTER \cite{Bast2007}} & Extended inverted index – inverted index with scores for each word ; combines prefix search and join operations & Proprietary module \\ \hline
\textbf{QAST \cite{Jitkrittum2009} } & Inverted index with term weighting ((Minimal Span Weighting)) & Lucene \\ \hline
\textbf{FREyA \cite{damljanovic2012freya}} & *N/A & Sesame/OWLIM (aka GraphDB)) \\ \hline
\textbf{QAKIS \cite{cabrio2012qakis}} & *N/A & *N/A \\ \hline
\textbf{MEANS \cite{BenAbacha2015}} & Inverted index & Terrier \\ \hline
\textbf{Watson/DeepQA \cite{kalyanpur2012structured,ferrucci2010building}} & Persistent disk caching & Watson Explorer Engine XML (VXML) \\ \hline
\end{tabular}
}
\caption{Comparison of the indexing platforms and data structures used by different QA systems. *N/A = no data available}
\label{comp_table_indexing}
\end{table*}

Table~\ref{comp_table_indexing} is an excerpt from a table in our exhaustive survey of open \textbf{QA systems}\footnote{The full data collection can be found at \url{https://goo.gl/FM1LM9}}.
Our current work in progress is focusing on benchmarking different datasets such as Wikidata~\cite{vrandevcic2014wikidata}, DBpedia~\cite{DBLP:conf/semweb/AuerBKLCI07}, and FreeBase~\cite{DBLP:conf/aaai/BollackerCT07} against a wide spectrum of indexing library platforms, such as Indri\footnote{\url{http://www.lemurproject.org/indri.php}}, Solr\footnote{\url{http://lucene.apache.org/solr/}}, ElasticSearch\footnote{\url{https://www.elastic.co/products/elasticsearch}}, Sphinx\footnote{\url{http://sphinxsearch.com/}}, Neo4j\footnote{\url{http://neo4j.com/}}, Titan\footnote{\url{http://thinkaurelius.github.io/titan/}}, Xapian\footnote{\url{http://xapian.org/}}, and Terrier\footnote{\url{http://www.terrier.org/}}.

\subsubsection{Data Quality Challenge}
Recent advancements in the fields of Web of Data and Data Science have led to an outburst of standards related to structured data\footnote{The amount not only of structured, but also of semi-structured and unstructured data available online is also steadily increasing; however, for the purpose of our work we assume that such data has first been translated to the RDF data model using standard tools, e.g. from the Linked Data Stack~\cite{aue+11}.} such as RDF(a), Linked Data, schema.org, etc., to an increasing amount of such data, and to a wide range of tools to produce, manage and consume such data.
To be available for ready consumption, especially in open question answering systems, any such data sources should meet a certain level of quality, e.g., defined by benchmarks.
Quality can generally be defined as ``fitness for use'', but there are a lot of concrete factors that influence a dataset's fitness for use in question answering\footnote{In this section, we do not abbreviate ``question answering'' as ``QA'' to avoid confusion with ``quality assessment''.} settings and in specific application domains.
Recently, a number of research activities have been concerned with automating the assessment of linked data quality.
Debattista, who has developed one such tool (Luzzu~\cite{DebattistaEtAl:LuzzuFramework2015}), provides an overview of other state-of-the-art tools~\cite{DebattistaEtAl:LuzzuFramework2015}, including one by Flemming~\cite{Flemming2008}, as well as Sieve~\cite{mendes2012sieve}, RDFUnit~\cite{kontokostas2014test}, TripleCheckMate~\cite{zaveri2013user}, LinkQA~\cite{gueret2012assessing}, and LiQuate~\cite{ruckhaus2013analyzing}.
In this section, we summarize the concrete criteria by which the quality of linked data can be assessed, with a special focus on those criteria that are relevant to question answering.

In a comprehensive review of literature and systems, Savors et al.~\cite{Zaveri2012:LODQ} have identified the dimensions of linked data quality and categorized them as follows:

\begin{itemize}
\item \textbf{Accessibility dimensions}: This category covers aspects related to retrieving and accessing data, which includes full or partial access and different technical means of access (e.g. the possibility to download a data dump vs. the availability of a SPARQL endpoint, i.e. a standardized query interface).
\begin{itemize}
\item Availability is generally defined as the ease of access with which particular information is obtainable or rapidly retrievable for readily consumption.
In a linked data context, availability can be referred to as the accessibility of a SPARQL endpoint or RDF dumps or dereferenceable URIs.
\item Interlinking is relevant as it refers to the data integration and interoperability.
The output of interlinking is a \emph{linkset}, i.e. a set of RDF triples linking subjects and recognized related objects.
\item Security denotes the degree to which a particular dataset is resistant to misuse or alteration without appropriate user access rights. 
\item Verifiability, usually by an unbiased third party, addresses the authenticity and correctness of the dataset.
Verifiability is typically enabled by provenance metadata.
\end{itemize}
\item \textbf{Intrinsic dimensions}: This category covers aspects that are independent of the user's context, or the out of the application context – such as accuracy and consistency.
\begin{itemize}
\item Accuracy refers to the degree of a dataset correctly representing the captured real world facts and figures in the form of information with high precision. 
\item Consistency refers to the independence from logical, formal or representational contradictions of a dataset with respect to others.
\item Completeness is referred to as the degree to which information in the dataset is complete or not missing.
The dataset should have all the required objects or values for a given task in order to be considered as complete.
Thus, arguing intuitively, completeness is one of the concrete metrics for linked data quality assessment. 
\end{itemize}
\item \textbf{Contextual dimensions}: This category is concerned with the context of the task being pursued.
\begin{itemize}
\item Timeliness is concerned with the freshness of data over time or timeliness, i.e. the regularity of updates or merges and so on. 
\item Understandability can be achieved by providing appropriate human readable annotations to a dataset and its entities, and by consistently following a certain regular expression as a pattern for forming entity URIs.
\item Trustworthiness is concerned with the reliability or trustworthiness of the data and its source.
\end{itemize}
\item \textbf{Representational dimensions}: This category is concerned with the design and representation of the data and its schema.
For instance, understandability and interpretability.
\begin{itemize}
\item Interpretability refers to adhering to the standard practice of representing information using appropriate notations, symbols, units and languages.
\end{itemize}
\end{itemize}

Data quality dimensions in all of these categories can be relevant in question answering scenarios.
In a preliminary study~\cite{thakkar2016linked}, we evaluated few selected metrics mentioned above on two popular datasets of linked data namely, Wikidata and DBpedia\footnote{Freebase used to be another popular cross-domain dataset but support for it has expired, which is why we did not consider it; cf.\ \url{https://www.freebase.com/}.}.
We evaluated the subsections extracted from these datasets on categories such as ``politicians'', ``Movies'', ``Restaurants'' and ``Soccer players''.
The slices have been made available to the common public and can be found: \url{https://goo.gl/Kn6Fom} (DBpedia slices), \url{https://goo.gl/5aTkLp} (Wikidata slices).
The detailed information regarding the data slice statistics can be found from the work~\cite{thakkar2016linked} which is selected from a detailed report made public at \url{https://goo.gl/ignzzI}.
For this evaluation we have obtained four slices of both DBpedia and Wikidata, namely \textit{Restaurants, Politicians, Films} and \textit{Soccer players}.
From this preliminary study, we could so far observe that Wikidata dominates DBpedia for the considered setting.
The detailed scores and discussion can be found from the mentioned work and the spreadsheet.

Our next step is \textit{(i)} implement and evaluated the pending metric from the above work and \textit{(ii)} to identify more systematically what other dimensions and metrics of data quality are specifically relevant in the typical application domains of question answering, or sufficient for determining a dataset's ``fitness'' for question answering.
Having identified such dimensions, we have two goals: \textit{(a)} identifying datasets that are suitable for question answering at all, and \textit{(b)} evaluate these metrics on a major part of the LOD Cloud\footnote{LOD Cloud: \url{http://lod-cloud.net/}} datasets, identifying more specifically what quality problems they still suffer from.

Regarding implementation, in our recent study~\cite{thakkar2016linked}, we evaluated the results on DBpedia and Wikidata slices using the metrics that the Luzzu linked data quality assessment framework already provides. We look forward to extending Luzzu to use in a question answering setting, that further existing implementations of metrics in Luzzu can be specifically adapted to make them suitable for quality assessment related to question answering, and that, finally, Luzzu's flexible extensibility even enables us to implement new metrics that may be required.
In summary, our near-future work will be concerned with defining a generally and flexibly applicable framework for automating the process of rigorously assessing the quality of linked datasets for question answering by identifying, formalizing and implementing the required metrics. 

\subsubsection{Distributed Heterogeneous Datasets}

The decentralized architecture of the Web has produced a wealth of knowledge distributed across different data sources and different data types. 
Question answering systems consume different types of data: \textit{structured}, \textit{semi-structured} or \textit{unstructured} data. 
Most question answering systems uses either of these types of data to answer user queries.
Only few systems exploit the wealth of data on the Web by combining these types of data.
Hybrid question answering systems are able to answer queries by combining both structured and unstructured types of data.
HAWK~\cite{Usbeck}, for instance, provides entity search for hybrid question answering using Linked Data and textual data.
HAWK is able to achieve an F-measure of up to 0.68 on the QALD-4 benchmark.

Most question answering systems today uses a single source to answer users question. 
It should rather be possible to answer questions imposed by a user by combining different interconnected sources.
The challenges imposed by the distributed nature of the Web are, on the one hand, finding the right sources that can answer user query and, on the other hand, integrating partial answers found from different sources.
Source selection is one of the challenges in federated question answering approaches. 
In~\cite{SHE+13}, the authors presented an approach to construct a federated query from user supplied (natural language) questions using disambiguated resources.

Answers may come from different sources which have different data quality and trust levels, ranking and fusion of data should be applied to select the best sources.



The amount of data to be used to answer users' queries should also be balanced with the response time.

\subsection{Interoperability Challenge}
The field of QA is so vast that the list of different QA systems can go long. Many Question Answering systems Based on specific domains have been developed. Domain-specific QA systems, for example \cite{DBLP:conf/ihi/AbachaZ12} are limited to a specific knowledge, for example medicine. They are known as \emph{closed domain} QA systems. However, when scope is limited to an explicit domain or ontology, there are less chances of ambiguity and high accuracy of answers. It is also difficult and costly to extend closed domain systems to a new domain or reusing it in implementing a new system.To overcome the limitations of closed domain QA systems, researchers have shifted their focus to \emph{open domain} QA systems.  FREyA \cite{DBLP:conf/esws/DamljanovicAC11}, QAKiS \cite{DBLP:conf/semweb/CabrioCAMLG12}, and  PowerAqua \cite{lopez2011poweraqua} are few examples of open domain QA systems which use publicly available semantic knowledge for example DBpedia \cite{DBLP:conf/semweb/AuerBKLCI07}.

While many of these system achieved significant performance for special use cases, a shortage was observed in all of them. We figured out that the existing QA systems suffer from the following drawbacks: (1) potential of reusing its components is very weak, (2) extension of the components is problematic, and (3) interoperability between the employed components are not systematically defined. There is little, but a work towards interoperable architecture, e.g. QA archiecture developed by OKBQA\footnote{Open Knowledge base and Question Answering (http://okbqa.org)}. Interoperability of different QA tools and components is required to enhance QA process which is still missing at the conceptual level and currently more focused on implementation details. Therefore, there is a need for a descriptive approach that define a conceptual view of QA systems. This approach must cover all needs of current QA systems and be abstracted from implementation details. Moreover it must be open such that it can be used in future QA systems. The generalized approach for architecture or ontology of a QA system and semantic search must focus to bring all state-of-the advancement of QA under a single umbrella \cite{singhtowards}. We envisioned that a generalized vocabulary for QA will be an abstraction level on top of all the existing QA approaches and will provide interoperability and exchangeability between them. This generalized vocabulary can be further used to integrate different components and web services within a QA system \cite{bothqanary}. 

\section{Conclusion and Future Roadmap}
In this paper, we presented an exhaustive overview of all the open challenges being still controversial for developing a question answering system.
The intuition is that Linked Data which provides advantages such as semantic metadata and interlinked dataset can influence all of the four major elements  (i.e. interface, parsing, data and component interoperability) which play a key role in Question Answering systems.
As our future research agenda, we are steering our research on all of the discussed issues with the focus of employing Linked Data technology to promote question answering capabilities.

\textbf{Acknowledgments} Parts of this work received funding from the European Union's Horizon 2020 research and innovation program under the Marie Sklodowska-Curie grant agreement No. 642795 (WDAqua project).

\bibliographystyle{plain}
\bibliography{bib/paper}
\end{document}